\documentclass[aps,twocolumn,
superscriptaddress,showpacs,showkeys,amsmath,amssymb,floatfix]{revtex4}
\usepackage{color}
\usepackage{graphicx}
\usepackage{epsfig}
\usepackage{dcolumn}
\usepackage{bm}
\usepackage{mathtools}
\usepackage{amssymb}
\usepackage{bmpsize}
\usepackage{calrsfs}
\DeclareMathAlphabet{\pazocal}{OMS}{zplm}{m}{n}
\usepackage{amsmath}
\usepackage{subfigure}
\newcommand{\be}{\begin{equation}}
\newcommand{\ee}{\end{equation}}
\newcommand{\bea}{\begin{eqnarray}}
\newcommand{\eea}{\end{eqnarray}}

\newcommand{\hu}{{\hat u}}
\newcommand{\hv}{{\hat v}}

\newcommand{\tA}{{\tilde A}}

\newcommand{\tC}{{\tilde C}}

\newcommand{\tT}{{\tilde T}}

\newcommand{\pro}{\partial}

\newcommand{\bfA}{{\mathbf A}}

\newcommand{\bpsi}{{\bar\psi}}

\newcommand{\ba}{\begin{array}}
\newcommand{\ea}{\end{array}}

\newcommand{\nn}{\nonumber}

\newcommand{\cL}{{\cal{L}}}

\begin{document}
\title{Weyl multiplet structure of QCD}
\author{Dmitriy G. Pak}
\affiliation{School of Physics and Astronomy,
 Sun Yat-sen University, Zhuhai, 519000, China}
\affiliation{Bogolyubov Laboratory of Theoretical Physics, JINR, 141980 Dubna, Russia}
\affiliation{Physical-Technical Institute of Uzbek Academy of Sciences, Tashkent 100084, Uzbekistan}
\author{Pengming Zhang}
\affiliation{School of Physics and Astronomy,
 Sun Yat-sen University, Zhuhai, 519000, China}
 \author{Takuya Tsukioka}
\affiliation{School of Education, Bukkyo University, Kyoto 603-8301, Japan}
\begin{abstract}
We consider Weyl group symmetric structure of $SU(3)$ quantum chromodynamics (QCD) with one flavor quark.
It has been demonstrated that the Weyl group as a finite color subgroup of $SU(3)$ provides an intrinsic 
color symmetry of quark and gluon fields on the mass shell.
We show that the standard QCD equations of motion lead to systems of equations for the quark and gluon
fields forming non-trivial irreducible multiplets in one- and two-dimensional irreducible representations of the Weyl group.
Some implications of Weyl multiplet structure in QCD and hadron physics are considered.
\end{abstract}
\maketitle

\subsection {Introduction}

  A modern theory of strong interactions is represented by quantum chromodynamics (QCD) which
is a quantum theory of the classical $SU(3)$ Yang-Mills theory \cite{yang-mills} with quarks.
Despite on tremendous progress in QCD for passed fifty years since its invention, especially in hadron
phenomenology, the construction of
a rigorous non-perturbative theory of QCD, or a strict quantum Yang-Mills theory, still represents
one of unresolved Millennium problems announced by Clay Institute of Mathematics \cite{Mill2000}.
The difficulties of constructing a rigorous quantum Yang-Mills theory arise already at the first steps of introducing
the concepts of fundamental particles, quarks and gluons, as well as in the subsequent definition of the Hilbert space
of quantum states including a non-trivial vacuum.
One of principal problems is related to a known mathematical fact, that the special unitary group $SU(3)$, and its
continuous subgroups, do not admit non-trivial one-particle singlet representations.
  In electro-weak theory the absence of such non-trivial
singlet representations does not cause any problems in defining the
vector bosons as elementary particles, since the
electroweak symmetry  $SU(2)\times U(1)$ is spontaneously broken.
Contrary to this, in QCD, in the confinement phase, the color symmetry is preserved.
Thus, to determine the fundamental particles in QCD, it is necessary first to construct classical single-particle
dynamical quark and gluon solutions which form invariant subspaces under color transformations.

In 1980s Wilson, Kogut and Susskind made an important observation, that in the confinement phase
the vacuum must be non-degenerate and color invariant \cite{wilson1974, kogut1975,polyakov77}.
Another important observation was made by `t Hooft who conjectured that color
 confinement assumes the vacuum to be color neutral and, as a consequence, the vacuum structure of QCD
 should be described by an Abelian projected QCD \cite{thooft81}.
This implies that one has to find color invariant vacuum gluon solutions.
 A serious obstacle in solving this problem was the quantum instability of known vacuum gluon solutions
 \cite{savv, N-O}. The problem has been resolved in \cite{plb2018},
 where stable regular gluon solutions have been found in the space of stationary field configurations.
It turns out that such solutions possess remarkable features: the solutions
are symmetric under Weyl group transformations, and they describe
one-particle gluons which reveal their own intrinsic color symmetry \cite{plb2023}.

 Certainly, it was known before, that Weyl group symmetry describes color invariant properties of hadrons.
Various Weyl symmetric QCD models have been widely considered in the past: phenomenological
Ginsburg-Landau models \cite{KomaToki2000,KomaToki2001,TokiSuganuma2000},
models with Wu-Yang type monopoles \cite{cho1982,godd1977},
instantons and topological knot solutions\cite{pak2012}, and some
supersymmetric field theories \cite{supsymm}.
In these studies, only composite singlets containing multiple particles were considered.
Due to the lack of nontrivial singlet representations of the color group $SU(3)$, it might seem that
the existence of singlet quarks is incompatible with exact color symmetry.
However, the found single-particle non-Abelian Weyl symmetric gluon solutions
in pure gluodynamics \cite{plb2018} indicate the possible existence of singlet quark
solutions with non-trivial color properties that are hidden when considering off-shell quark-gluon
fields.

 In the present paper we consider the Weyl multiplet structure of quark and gluon fields on mass shell in $SU(3)$
QCD with one color quark triplet in fundamental representation.
We show that equations of motion of standard QCD
 admit solutions for constituent quarks which belong to standard $N$-dimensional  ($N=1,2,3$)
 vector representations of the Weyl group including a singlet representation which
 describes a constituent single quark dressed in a singlet gluon field.
 Some implications of non-trivial Weyl multiplet structure of quarks and gluons are considered.

 \subsection{Weyl symmetry of classical QCD Lagrangian}

We start with main definitions of the conventional $SU(3)$ QCD
with one flavor quark.
The original $SU(3)$ QCD Lagrangian is defined in a standard form
\bea
 {\cal L}_{\rm QCD}&=&{\cal L}_{\rm YM}+{\cal L}_{\rm q}, \label{LagrQCD} \\
{\cal L}_{\rm YM}&=& -\frac{1}{4} F_{\mu\nu}^a F^{a\mu\nu}, \label{LagrYM} \\
{\cal L}_{\rm q}&=& \bar {\hat \Psi} \Big[ i \gamma^\mu (\pro_\mu -i g \vec \bfA_\mu)
-m\Big]{\hat \Psi}, \label{Lagrq} \\
&&\vec \bfA_\mu \equiv A_\mu^a T^a, \nn
\eea
where ${\cal L}_{\rm YM}$ is $SU(3)$ Yang-Mills Lagrangian, ${\cal L}_{\rm q}$ is a
Lagrangian for $SU(3)$ quark triplet ${\hat \Psi}=(\psi^1,\psi^2,\psi^3)$ defined
in fundamental representation, the quark component fields $\psi^p(x)$ are described by
Dirac spinor fields (index values $(p=1,2,3)$ denote $I,U,V$-type fields),
and $F^a_{\mu\nu}$ is a field strength defined in terms of the gluon vector potential $A_\mu^a$ $(a=1,2,...,8)$.
 The full QCD Lagrangian implies the following equations of motion
 \bea
&&(D^\mu F_{\mu\nu})^a =-\dfrac{g}{2}  \bar {\hat \Psi} \gamma_\nu \lambda^a{\hat \Psi}\equiv-J_\nu^a,
                                                                                                                        \label{eqA}\\
&& \Big[ i \gamma^\mu (\pro_\mu -i g A_\mu^a T^a)-m\Big]{\hat \Psi}=0, \label{eqPsi}
\eea\\
where $D_\mu$ is a covariant derivative in adjoint representation, and $J_\mu^a$ is
the quark current.

Let us consider a general Weyl symmetric structure of QCD.
Generators $T^a$ of $SU(3)$ fundamental representation are expressed in terms of
Gell-Mann matrices, $T^a=\dfrac{1}{2} \lambda^a$.
The color group $SU(3)$ contains three $I,U,V$-type subgroups $SU(2)$ generated by
 the following generators
\bea
I:&& \{T^1,T^2,T^3\} ;\nn\\
U: &&\{T^4,T^5, \dfrac{1}{2} (-T^3+\sqrt 3 T^8\} ;\nn\\
V: && \{T^6,T^7, \dfrac{1}{2} (-T^3-\sqrt 3 T^8\} . \label{IUVsect}
\eea
The explicit Weyl symmetric structure is manifested in
the Cartan-Weyl basis of $SU(3)$ Lie algebra which contains
two generators $T^3, T^8$ of the Cartan subalgebra and three complex off-diagonal $I,U,V$-type
generators $T^p_\pm$
\bea
T^I_\pm&=&\dfrac{1}{2}(T^1\pm i T^2), \nn \\
T^U_\pm&=&\dfrac{1}{2}(T^4\mp i T^5), \nn \\
T^V_\pm&=&\dfrac{1}{2}(T^6 \pm i T^7). \label{IUVgen}
\eea
Root vectors $r^p_\alpha$ ($\alpha=(3,8)$) are defined by six eigenvalues of 2-component
operator $T^\alpha =(T^3, T^8)$ acting on $SU(3)$ Lie algebra by commutators
\bea
[T^\alpha, T^p_\pm]= r^p_\alpha T^p_\pm
\eea
It is sufficient to use three root vectors which form a right star in the Cartan plane
$(T^3,T^8)$:
 \bea
&&r^1_\alpha=(1,0), \nn\\
&& r^2_\alpha=(-1/2,\sqrt 3/2), \nn\\
&& r^3_\alpha=(-1/2,-\sqrt 3/2).  \label{roots}
 \eea
  The remaining three roots are defined by reflection $r^p_\alpha \rightarrow -r^p_\alpha$.
The symmetry group of the root system is an example of a non-vector representation
of the Weyl group, which is useful in some applications in non-linear theories that do not allow
a linear vector structure of the solution space \cite{plb2023}.
  The Weyl group $W(SU(3))$ is isomorphic to the symmetric group $S_3$ acting on the root vectors
 by permutations (main definitions of Weyl group representations are given in
 Supplemental Material to \cite{plb2023}). Decomposition of the gauge potential
  $\vec \bfA_\mu$ in the Cartan-Weyl basis leads
 to introducing complex field variables for off-diagonal gluons
\bea
W_\mu^1=\dfrac{1}{\sqrt 2}(A^1_\mu+i A_\mu^2),\nn\\
W_\mu^2=\dfrac{1}{\sqrt 2}(A^4_\mu-i A_\mu^5),\nn\\
W_\mu^3=\dfrac{1}{\sqrt 2}(A^6_\mu+i A_\mu^7).  \label{complexW}
\eea
It is known, that classical Lagrangians ${\cal L}_{YM}$ and ${\cal L}_q$ can be written in explicit
Weyl symmetric form \cite{flyvb}
\bea
{\cal L}_{YM}&=&\sum_{p=I,U,V} \Big\{ -\dfrac{1}{6} (F^p_{\mu\nu})^2 + \dfrac{1}{2}\big| D^p_{\mu} W_\nu^p-
 D^p_{\nu} W_\mu^p \big|^2 \nn \\
&& -ig F_{\mu\nu}^p W^{*p}_{~\mu} W_\nu^p
 -\dfrac{9}{4}g^2 \big[(W^{*p\mu} W_\mu^p)^2\nn \\
 &&-(W^{*p\mu}W^{*p}_\mu)^2 (W^{p\nu} W^p_\nu)^2\big]\Big\}, \label{LYM} \\
 \cL_q&=& \sum_p \bpsi^p \Big (i \gamma^\mu \pro_\mu -m +
 \dfrac{g}{2} \
^\mu (A_\mu^p)_{(w)}\Big )\psi^p, \label{lagrqp}\\
&&F^p_{\mu\nu}=\pro_\mu (A^p_\nu)_r-\pro_\nu (A^p_\mu)_r,  \\
&&(A^p_\mu)_r=A^\alpha_\mu r^p_\alpha, ~~~~~(A^p_\mu)_w=A_\mu^\alpha w^{p\alpha}, \label{Fp}
 \eea
where $D^p_{\mu}W^p_\nu=(\pro_\mu -ig (A^p_\mu)_r)W_\nu^p$ is $U(1)$ covariant derivative,
and indices ``r,w'' indicate that $p$-components $(A^p_\mu)_{r,w}$ of the Abelian gluon fields $A^{3,8}_\mu$
are defined with using roots or weights, respectively. The Weyl group transformations on a
space of quark-gluon fields are realized by all possible permutations of p-components of
quark and gluon fields from $I,U,V$-sectors.
The complex gluon fields $W^p_\mu$ and quark fields $\psi^p$ form invariant subspaces
corresponding to the standard three-dimensional
vector representation $\Gamma_3$ of the Weyl group.
The Weyl components of Abelian gluons, $(A^p_\mu)_r$ and $(A^p_\mu)_w$),
form a standard irreducible two-dimensional vector representation $\Gamma_2$.
Note, that vector representation $\Gamma_3$ is reducible and can be decomposed into a sum
of two irreducible representations $\Gamma_1$ and $\Gamma_2$.

It is known that  local gauge symmetry reflects a fundamental
gauge equivalence principle which determines the evolution of a physical system.
The local gauge symmetry can be fixed by removing all pure gauge field degrees of freedom from
the gluon fields.
After gauge fixing the space of arbitrary quark-gluon fields is reduced to a space
of gauge equivalent classes. Each equivalence class contains still a wide class of quark-gluon fields off-shell.
One can impose gauge fixing conditions and additional appropriate solution ansatz in such a way that
the space of quark-gluon solutions will be invariant only under transformations of the Weyl group
which is a finite color subgroup of the full color group $SU(3)$.
In that case, a space of Weyl symmetric solutions will be divided into invariant subspaces corresponding
to Weyl group multiplets forming representations $\Gamma_N$ ($N=1,2,3$). Our purpose is to study the
 Weyl multiplet structure of the space of quark-gluon solutions,
especially, to investigate whether quark-gluon solutions to exact QCD equations of motion
form invariant subspaces corresponding to irreducible Weyl group representations $\Gamma_1$ and $\Gamma_2$.
   \subsection{Weyl symmetry of the quantum effective potential}

    We consider one-loop QCD effective potential with Abelian external color magnetic gluon field.
    The one-loop effective potential $V_{eff}$ describes the vacuum energy
    density as a function of vacuum gluon condensate parameters $F^2_p=(F_{\mu\nu}^p)^2$.
    In constant field approximation the analytical expression for one-loop effective potential in
    the Weyl symmetric form has been obtained in \cite{flyvb,mpla2006}
   \bea
   V_{eff}(F_p)=\sum_p \Big \{ \dfrac{F_p^2}{6}+\dfrac{11 g^2}{96\pi^2}
   F_p^2 (\ln \dfrac{g F_p}{\mu}-c) \Big \}, \label{Veff}
    \eea
    where $g$ is a coupling constant, $\mu$ is a mass parameter, and $c=1.292...$
    (in the modified minimal subtraction scheme \cite{mpla2006}).
Weyl components of the gluon field strength,  $F_p^2=(F_{\mu\nu}^p)^2$,
    are defined by Eqn. (\ref{Fp}) in terms of  corresponding gluon potentials $(A^p_\mu)_r$ (for brevity of notations
    we omit the index $r$ in this section).

    Note, that p-components $A_\mu^p$ satisfy an identity
\bea
    A^I_\mu+A^U_\mu+A^V_\mu=0,
\eea
which implies that $A_\mu^p$ belong to a two-dimensional
representation $\Gamma_2$.
The corresponding field strengths $F_{\mu\nu}^p$ form a two-dimensional representation
$\Gamma_2$ as well. Since representation $\Gamma_2$ is irreducible, one can not reduce it
to the singlet representation $\Gamma_1$ containing non-zero gluon fields.
The  imposing the reduction constraints
\bea
A^I_\mu=A^U_\mu=A^V_\mu,
\eea
necessary for definition of the representation $\Gamma_1$, imply only a trivial singlet, $A_\mu^p=0$,
and vanishing the initial Abelian gluon potentials $A_\mu^{3,8}$.
Nevertheless, in the space of gluon field strengths one can define a non-zero singlet gluon field
by imposing scalar constraints on the squared  field strengths
\bea
F_I^2=F_U^2=F_V^2. \label{Fpsqr}
\eea
Note that constant Abelian gluon fields are determined by three independent parameters,
either by $F_p^2$, or by two field strength parameters
 $F_3^2= (F_{\mu\nu}^3)^2, F_8^2=(F_{\mu\nu}^8)^2$, and one angle parameter
 $\cos \beta=(F_{\mu\nu}^3 \cdot F_{\mu\nu}^8)/F_3 F_8$.

 It has been shown by Flyvberg \cite{flyvb}, that effective potential (\ref{Veff}) has three minimums:
two degenerate local minimums  in the plane $\{F_3,F_8 \}$ at the points (i) $(F_3=H_0, F^8=0)$ and
(ii) $(F_3=1/2 H_0, F_8=\sqrt{3}/2 H_0)$, in a case of parallel vector magnetic fields
$\vec F_3, \vec F_8$; and the third deepest absolute minimum at the point (iii) $F_3=F_8=2^{-1/3}H_0$ in a case of
orthogonal vector gluon fields, $\cos \beta=0$. Taking into account that effective potential is invariant under reflections
$F^{3,8}\rightarrow \pm F^{3,8}$, one obtains six degenerate minimums forming a Weyl sextet in two-dimensional
representation $\Gamma_2$,
 and one non-degenerate absolute minimum corresponding to the Weyl singlet representation $\Gamma_1$.
The degenerate vacuums represent false vacuums, this becomes evident if we consider the effective potential
$V_{eff}$ as a function of three Weyl invariant variables $F^p$. In these variables, the effective potential has only
one absolute minimum at the symmetric point $F_I=F_U=F_V$.

Note, that Weyl invariants $F_p$ are gauge invariant under full $SU(3)$ transformations,
so, they can be expressed in terms of three Casimir invariants $C_2, C_4, C_6$
as follows \cite{nayak,mpla2006}
\bea
F_p^2=\dfrac{\sqrt{3 C_4}}{2} \sin \phi_p +C_2,
\eea
where $\phi_p$ are three basic solutions of the equation
\bea
\sin 3 \phi= \dfrac{2}{\sqrt 3} \dfrac{(8C_6-C_2 C_4)}{C_4^{3/2}},
\eea
and Casimir invariants are defined in a standard manner
\bea
&&C_2=\dfrac{1}{2} (F_{\mu\nu}^a)^2, \nn\\
&&C_4=(d^{abc}F_{\mu\nu}^b F_{\mu\nu}^c)^2, \nn \\
&& C_6=(d^{abc}F_\mu\nu}^a F_{\nu\rho^b F_{\rho \sigma}^c)^2.
\eea
Irreducible Weyl group representations $\Gamma_2$ and $\Gamma_1$ are classified by Casimir invariants
as follows \cite{mpla2006}:
three positive Weyl invariants $F_p$ forming the multiplet  $\Gamma_2$ are defined by
the following Casimir invariants
\bea
&&C_2=H_0^2,\nn\\
&&C_4=\frac{4}{3} H_0^4,\nn\\
&&C_6=0,
\eea
which imply a constraint $\cos\beta =\pm 1$, i.e., the magnetic fields are parallel.
The singlet representation $\Gamma_1$ corresponding to the absolute minimum of the effective potential
 is determined by the following Casimir invariants
\bea
&&C_2=H_0^2, \nn \\
&&C_4=C_6=0,
\eea
which imply that magnetic fields  $\vec F_3, \vec F_8$ are orthogonal.

The structure of one-loop effective potential implies
that Weyl symmetric solutions forming Weyl multiplet $\Gamma_2$
have a higher symmetry than of-shell gluon fields characterized, in general, by all non-vanishing Weyl invariants
$F_p$, whereas the singlet gluon solution is defined by equal Weyl invariants $F_p$, or by only one non-zero Casimir
invariant $C_2$. So that, the singlet gluon solution reveals the
highest inherent color symmetry of the gluon itself. Therefore, the Weyl group color symmetry leads to
the existence of unique non-degenerate color invariant vacuum which provides an exact color symmetry,
and, as a consequence, the existence of the color confinement phenomenon in agreement with
the observation of Wilson, Kogut and Susskind  \cite{wilson1974, kogut1975}.

Certainly, the existence of the singlet structure in the space of constant gluon fields is
not merely satisfactory, since in the full QCD with quarks
 the main quantum gluon field variable is the gluon vector potential which enters the
quark-gluon interaction terms. Besides, the constant gluon fields possess
the quantum instability \cite{N-O}. It has been proved, that quantum stable
vacuum gluon condensate can be described by stationary gluon solutions \cite{plb2018}.
Therefore, in the QCD with quark matter one needs to prove the existence of
a singlet non-constant vector gluon potential.

\subsection{Abelian projection consistent with singlet quark structure}

 Following the idea of the dominant role of the Abelian projection in
of QCD in the confinement phase \cite{thooft81} we will find a proper
Abelian projection that is suitable to reveal the Weyl multiplet structure of quark-gluon solutions.
The problem is that an Abelian projection can be defined in numerous ways.
In the conventional QCD a simple Abelian projection is defined by two Abelian gluon fields
$A_\mu^{3,8}$ corresponding to the generators of Cartan subalgebra  $T^{3,8}=\dfrac{1}{2} \lambda^{3,8}$.
The generators have three common eigenvectors $\hu^p$  with respective eigenvalues
equaled to weights $w^{p\alpha}$
\bea
&& \hu^1=(1,0,0), ~~\hu^2=(0,1,0), ~~\hu^3=(0,0,1), \label{eigenvecT38} \\
&& T^{\alpha} \hu^p=w^{p\alpha}\hu^{p}. \nn
\eea
One can decompose the quark fundamental triplet $\hat \Psi$ in the color vector basis $\{\hu^p\}$
\bea
\hat\Psi=(\psi^1,\psi^2,\psi^3)=\psi^1 \hu^1+\psi^2 \hu^2+\psi^3 \hu^3, \label{qdec}
\eea
where $\psi^p$ are Dirac spinor fields describing three color quarks,  (``red'', ``blue'', ``yellow''),
with color charges defined by weights  $w^{p\alpha}$.
Substitution of the decomposition (\ref{qdec}) into the quark Lagrangian $\cL_q$, (\ref{Lagrq}),
leads to Weyl symmetric form of the quark Lagrangian $\cL_q$  which is factorized into a direct
sum of three $I,U,V$-terms (\ref{lagrqp}). The quark triplet $\hat\Psi$ belongs to a three-dimensional
vector representation $\Gamma_3$ of the Weyl group which is reducible. So
one can extract a Weyl singlet from the quark triplet $\Psi$ by identifying all the components
of the quark triplet as follows
\bea
\psi^1=\psi^2=\psi^3 \equiv \psi^s. \label{psi123}
\eea
With this, the decomposition (\ref{qdec}) provides a formal ansatz for the singlet quark solution $\psi^s$
\bea
\hat \Psi^s= (\psi^s,\psi^s,\psi^s) \equiv \psi^s \hu^s, \label{singans1}
\eea
where the color vector $\hu^s=(1,1,1)$ forms a vector basis in the representation space of the singlet
Weyl group representation $\Gamma_1$.
The main question remains,
whether ansatz (\ref{singans1}) leads to a non-trivial physical singlet quark solution.
One can verify that substitution of the ansatz (\ref{singans1}) into Eqn. (\ref{eqA}) leads to
vanishing of the quark current $J_\mu^a$
\bea
&&J^a =0, \label{currents38}
\eea
 and Eqn. (\ref{eqPsi}) turns into a free equation for a non-interacting colorless quark which is unphysical.
  It is clear, that inconsistency of the simple Abelian projection with the quark singlet structure
 is a result of a particular choice of the Cartan generators $T^3, T^8$ which do not admit a
 color singlet vector $\hu^s$ as an eigenvector.
To construct a proper Abelian projection we start with all non-vanishing gluon fields $A_\mu^a$
defined in the standard basis of $SU(3)$ generators defined in terms of Gell-Mann matrices.
Note, that two sets of gluon fields $(A^1_\mu, A^4_\mu, A^6_\mu)$ and $(A^2, -A^5, A^7)$ realize
two three-dimensional Weyl representations $\Gamma_3$, as it follows from the Weyl
symmetric structure of the classical Yang-Mills Lagrangian written in the Cartan-Weyl basis,
(\ref{IUVsect}, \ref{IUVgen}, \ref{complexW}).
The representation $\Gamma_3$ is reducible, and by imposing constraints
\bea
&&A^1_\mu=A^4_\mu=A^6_\mu\equiv \tC_{8\mu} , \nn \\
&&A^2_\mu=-A^5_\mu=A^7_\mu\equiv \tC_{3\mu}, \nn \\
&& A_\mu^3=A_\mu^8\equiv 0, \label{A2singlet}
\eea
one can extract two singlet Weyl representations $\Gamma_1$ defined on the corresponding
one-dimensional vector spaces with the basis vectors $\tC_{8\mu}(1,1,1)$ and $\tC_{3\mu}(1,1,1)$.
Under the constraints (\ref{A2singlet}), the Lie algebra valued gluon potential takes
the following form
\bea
A_\mu^a T^a= \tC_{8\mu} \dfrac{1}{2} L+\tC_{3\mu} \dfrac{1}{2} Q,
\eea
where $L$ and $Q$ represent two color charge matrices defined as follows
\bea
  L=\lambda^1+\lambda^4+\lambda^6=\begin{pmatrix}
   0&1&1\\
  1&0&1\\
   1&1&0\\
 \end{pmatrix},
\qquad \nn \\
 Q=\lambda^2-\lambda^5+\lambda^7=\begin{pmatrix}
   0&-i & i\\
  i &0&-i\\
   -i&i&0\\
 \end{pmatrix}.
\label{LQ}
 \eea
 One can verify that charge matrices $L,Q$ commute to each other,
\bea
[L,Q]=0,
\eea
and form a new basis of a different Cartan subalgebra. Thus, two Abelian gluon fields $\tC_{3\mu}, \tC_{8 \mu}$
define a new Abelian projection which admits two standard singlet one-dimensional vector representations $\Gamma_1$
of the Weyl group. 

The commuting charge matrices $L, Q$ have three common eigenvectors $\hv^p=\{\hv^0, \hv^+, \hv^-\}$
with corresponding eigenvalues
\bea
&&\hv^0= (1,1,1),~~~~~~~~
\nn \\
&& \hv^\pm=(-\dfrac{1}{2}\pm\dfrac{i \sqrt 3}{2},-\dfrac{1}{2}\mp\dfrac{i \sqrt 3}{2},1), \label{upm}\\
&&L\hv^0=2 \hv^0, ~~~~~~~~~~Q\hv^0=0,  \nn\\
&&L\hv^\pm=-\hv^\pm, ~~~~~~~~Q\hv^\pm=\pm \sqrt 3 \hv^\pm, \nn \\
&&\hv^+\hv^-=\hv^-\hv^+=3,~~~\hv^+ \hv^+=0,~~~\hv^- \hv^-=0, \label{upmprops}\\
&& (\hv^+)^*=\hv^-.~~~~~~\nn
\eea
An important feature of matrices $(L,Q)$ is that eigenvector $\hv^0$ realizes the irreducible singlet
representation $\Gamma_1$,
and color vectors $\hv^\pm$ realize the two-dimensional irreducible representation $\Gamma_2$
due to the defining relationship
\bea
\hv_1^\pm+\hv_2^\pm+\hv_3^\pm=0.
\eea
The $SU(3)$ quark triplet ${\hat \Psi}$ can be decomposed in the basis of three
color vectors $\hv^p=\{\hv^0,\hv^+,\hv^-\}$
\bea
{\hat \Psi}=\psi^1 \hv^0+\psi^2 \hv^+ +\psi^3 \hv^-.  \label{qdecomp2}
\eea
The decomposition (\ref{qdecomp2}) provides
 decomposition of three-dimensional representation $\Gamma_3$ into a sum of
 two irreducible representations $\Gamma_1$ and $\Gamma_2$ containing the singlet quark $\psi^1$ and
 a quark doublet $\{\psi^+, \psi^-\}$ respectively.
 After substitution of this decomposition into original QCD equations of motion
(\ref{eqA}, \ref{eqPsi}) one obtains the following equations in terms of field components
\bea
&&\pro^\mu F_{\mu\nu}[\tC_{3\mu}]=-\dfrac{g\sqrt 3}{2}(\bar \psi^2\gamma_\nu\psi^2
                   -\bar\psi^3\gamma_\nu\psi^3),\nn \\
&&\pro^\mu F_{\mu\nu}[\tC_{8\mu}]=-g \bar \psi^1 \gamma_\nu \psi^1+
      \dfrac{g}{2}(\bar \psi^2\gamma_\nu\psi^2 +\bar\psi^3\gamma_\nu\psi^3), \nn \\
&&\big(i \gamma^\mu \pro_\mu -m +g\gamma^\mu \tC_{8\mu}\big) \psi^1=0, \nn \\
&&\big(i \gamma^\mu \pro_\mu -m -\dfrac{g}{2}\gamma^\mu \tC_{8\mu}
                  +\dfrac{g\sqrt 3}{2}\gamma^\mu\tC_{3\mu}\big) \psi^2=0, \nn \\
&&\big(i \gamma^\mu \pro_\mu -m -\dfrac{g}{2}\gamma^\mu \tC_{8\mu}
                   -\dfrac{g\sqrt 3}{2}\gamma^\mu\tC_{3\mu}\big) \psi^3=0. \label{psys1}
\eea
In a general, the system of Eqs. (\ref{psys1}) describes three quarks interacting with
two Abelian gluon fields. By imposing a simple constraint
\bea
\psi^1=0,
\eea
the equations (\ref{psys1}) reduce to a system of equations for two coupled constituent quarks
 $\psi^{2,3}$ dressed in  gluon fields $\tC_{3\mu}, \tC_{8\mu}$. The reduced system of equations 
 describes quark-gluon solutions which belong to the standard two-dimensional irreducible representation $\Gamma_2$.
The equations for singlet quark $\psi^1$ are obtained from equations (\ref{psys1}) by applying the
following reduction constraints
\bea
&&  \psi^2=\psi^3\equiv 0. \label{anspsi1}
\eea
As a result, one obtains a simple system of equations for a free singlet gluon field $\tC_{3\mu}$
and for the singlet constituent quark $\psi^1$ dressed in gluon field $\tC_{8\mu}$
\bea
&&\pro^\mu F_{\mu\nu}[\tC_{3\mu}]=0, \label{eqC3} \\
&&\pro^\mu F_{\mu\nu}[\tC_{8\mu}]=-g \bar \psi^1 \gamma_\nu \psi^1,  \label{eqC8} \\
&&\big(i \gamma^\mu \pro_\mu -m + g \tC_{8\mu}\big) \psi^1=0. \label{psys2}
\eea

We conclude,{\it  that in Abelian projected QCD a space of quark-gluon solutions for the quark triplet $\hat \Psi$ and
Abelian gluons $\tC_{3\mu}, \tC_{8\mu}$ contains two irreducible invariant subspaces,
one invariant subspace contains solutions for the singlet constituent quark dressed in gluon field $\tC_{8\mu}$,
 and another invariant subspace describes two constituent quarks
 from the two-dimensional representation $\Gamma_2$.} One should stress, that singlet quark $\psi^1$ and two
 singlet gluon fields $\tC_{3\mu}, \tC_{8\mu}$ originate from the corresponding reducible quark and gluon
 multiplets belonging to representation $\Gamma_3$ of the Weyl group which is a color subgroup of color group
  $SU(3)$. So that, the singlet quark and two singlet gluons
 possess an inherent color symmetry with respect to the Weyl group.
 Certainly, this intrinsic color symmetry corresponds to a small finite subgroup of $SU(3)$, but it is a real color
 symmetry in the old traditional sense \cite{weyl1952}, contrary to $SU(3)$ color symmetry
 which reflects a relational symmetry of quarks and gluons off-shell as members of $SU(3)$ multiplets.
Surely, the local gauge color symmetry keeps its basic prioritet in controlling the
dynamics of quarks and gluons, and selection of observable quantities which must be gauge invariant.

Note, that equations (\ref{eqC3}- \ref{psys2}) for the singlet constituent quark can be
generalized straightforward to a general case of $SU(N)$ QCD
\bea
&&\pro^\mu F_{\mu\nu}[\tC_{i\mu}]=0, \nn \label{eqA2N} \\
&&\pro^\mu F_{\mu\nu}[\tC_{r\mu}]=-\dfrac{g (N-1)}{2}\bar \psi^1 \gamma_\nu \psi^1, \nn  \label{eqA1N} \\
&&\big(i \gamma^\mu \pro_\mu -m + \dfrac{g(N-1)}{2} \tC_{r\mu}\big) \psi^1=0, \label{psys2N}
\eea
where $\tC_{i\mu}$ are Abelian gluon fields corresponding to Cartan generators $H_i$, ($i=1,...,r-1$),
``r'' is a rank of the group $SU(N)$. Gluon solutions $\tC_i$ form an irreducible $(r-1)$-dimensional Weyl multiplet
$\Gamma_{r-1}$.
One has only one constituent singlet quark $\psi^1 \in \Gamma_1$ dressed in a singlet gluon $\tC_r$.
Other $N-1$ quarks $\psi^{k}$ ($k>1,..., N)$) form irreducible $(N-1)$-dimensional Weyl multiplet  $\Gamma_{N-1}$
which describe $(N-1)$ constituent quarks.
Details of generalization of the present results to the case of $SU(N)$ Yang-Mills theory with fermions
will be presented in a subsequent paper \cite{diez1}.

\subsection{Cartan-Weyl basis consistent with quark-gluon singlet structure}

In the previous section we have constructed Cartan generators $(L,Q)$ which define an Abelian projection
consistent with Weyl singlet structure of quark and gluon solutions. To formulate a full QCD
including off-diagonal gluon fields one should restore a full $SU(3)$ Lie algebra with a new
Cartan subalgebra generators $L,Q$ and verify the consistence of the resulting QCD Lagrangian
with the standard QCD Lagrangian (\ref{LagrQCD}). It is suitable to construct a full $SU(3)$ Lie algebra
in a new Cartan-Weyl basis which  can be found straightforward using the given Cartan subalgebra $(L,Q)$
and root vectors.

We introduce normalized generators ($H_3, H_8$) of Cartan subalgebra expressed in terms of charge matrices
$(L,Q)$
\bea
H_8&=&\dfrac{1}{2 \sqrt 3}L, \nn \\
H_3&=&\dfrac{1}{2 \sqrt 3}Q. \label{H3prim}.
\eea
The Cartan-Weyl basis is defined by two Abelian Cartan generators $(H_8,H_3)$ and by six off-diagonal generators
$T^p_\pm$ ($p=I,U,V$) corresponding to $I,U,V$-type $SU(2$) subalgebras.
Using known commutator relationships of all generators which
are defined uniquely by the root vectors $r^p_\alpha$, one can find the following explicit expressions for
generators $T^p_\pm$ corresponding to $I,U,V$-type subalgebras.\\
\indent
Type $I$ subalgebra:\\
 \bea
 &&[T^I_+,T^I_-]=H_3, \nn\\
 &&[H_3, T^I_\pm]=\pm T^I_\pm,\nn\\
  &&[H_8, T^I_\pm]=0,\label{LQI}
 \eea
 where
\bea
&&T^I_+=\dfrac{1}{3\sqrt 2}\begin{pmatrix}
   -\dfrac{1}{2}+\dfrac{i \sqrt 3}{2}&-\dfrac{1}{2}-\dfrac{i \sqrt 3}{2}& 1\\
  -\dfrac{1}{2}-\dfrac{i \sqrt 3}{2} &1&-\dfrac{1}{2}+\dfrac{i \sqrt 3}{2}\\
   1&-\dfrac{1}{2}+\dfrac{i \sqrt 3}{2}&-\dfrac{1}{2}-\dfrac{i \sqrt 3}{2}\\
 \end{pmatrix},\nn \\
 &&T^I_-=(T^I_+)^\dagger.
 \eea
\indent
 Type $U$ subalgebra of $SU(2)$ is defined as follows
  \bea
 &&[T^U_+,T^U_-]=\dfrac{1}{2}H_3+\dfrac{\sqrt 3}{2} H_8 \nn\\
  &&[H_3, T^U_\pm]=\pm \dfrac{1}{2}T^U_\pm,\nn\\
  &&[H_8, T^U_\pm]=\pm \dfrac{\sqrt 3}{2}  T^U_\pm,  \label{CWU}
 \eea
 where
   \bea
&&T^U_+=\dfrac{1}{3\sqrt 2}\begin{pmatrix}
   -\dfrac{1}{2}+\dfrac{i \sqrt 3}{2}&-\dfrac{1}{2}-\dfrac{i \sqrt 3}{2}& 1\\
    -\dfrac{1}{2}+\dfrac{i \sqrt 3}{2}&-\dfrac{1}{2}-\dfrac{i \sqrt 3}{2}& 1\\
     -\dfrac{1}{2}+\dfrac{i \sqrt 3}{2}&-\dfrac{1}{2}-\dfrac{i \sqrt 3}{2}& 1\\
 \end{pmatrix},\nn\\
&&T^U_-=(T^U_+)^\dagger.\nn\\
\eea
\indent
Type V subalgebra is defined by generators $T^V_\pm$ and $H_3, H_8$ as follows
\bea
 &&[T^V_+,T^V_-]=-\dfrac{1}{2}H_3+\dfrac{\sqrt 3}{2} H_8, \nn\\
 &&[H_3, T^V_\pm]=\mp \dfrac{1}{2}T^V_\pm,\nn\\
  &&[H_8, T^V_\pm]=\pm \dfrac{\sqrt 3}{2}  T^V_\pm, \label{CWV}
 \eea
 with
  \bea
&&T^V_+=\dfrac{1}{3 \sqrt 2}\begin{pmatrix}
    -\dfrac{1}{2}-\dfrac{i \sqrt 3}{2}& -\dfrac{1}{2}+\dfrac{i \sqrt 3}{2}& 1\\
    -\dfrac{1}{2}-\dfrac{i \sqrt 3}{2}& -\dfrac{1}{2}+\dfrac{i \sqrt 3}{2}& 1\\
    -\dfrac{1}{2}-\dfrac{i \sqrt 3}{2}& -\dfrac{1}{2}+\dfrac{i \sqrt 3}{2}& 1\\
     \end{pmatrix},\nn\\
 &&T^V_-=(T^V_+)^\dagger. \nn\\
 \eea

 All generators $H_{3,8}, T^p_\pm$ are normalized to 1/2 and form the orthonormal Cartan-Weyl basis,
 \bea
 && Tr (H_3 H_3)=Tr(H_8 H_8)=\dfrac{1}{2}, \nn\\
&& Tr (T^a_+ T^b_-)=\dfrac{1}{2}\delta^{ab}, \nn\\
&&Tr(H_3 H_8)=Tr(H_{3,8} T^a_\pm)=Tr[T^a_\pm T^b_\pm]=0.
 \eea
  The generators of the obtained Cartan-Weyl basis form  the same Lie algebra as generators
  $T^3, T^8, T^p_\pm$ of the Cartan-Weyl basis defined in terms of Gell-Mann matrices
  in Section 2.\\
  \indent One can write the gluon potential $A_\mu^a T^a$ in the new Cartan-Weyl basis in complex
  notations
  \bea
&&  \tA^a_\mu \tT^a=\tA^3_\mu H^3+\tA^8_\mu H^8+\tA^I_{+\mu}T^I_++\tA^U_{+\mu}T^U_++\nn \\
 &&~~~~~~~~ \tA^V_{+\mu}T^V_+ + \tA^I_{-\mu}T^I_- +\tA^U_{-\mu}T^U_- +\tA^V_{-\mu} T^V_-.  \label{Adecomp}
     \eea
 Comparing the last expression with the decomposition of the gauge potential
 in the basis of generators given in terms of Gell-Mann matrices,
  one obtains relationships between
new field variables $(\tA_{\alpha \mu}, \tA^{I,U,V}_{\pm\mu})$ and original gluon fields $A^a_\mu$
\bea
&& \tA^3_{\mu}=\dfrac{1}{\sqrt 3}(A^2_{\mu}-A^5_{\mu}+A^7_{\mu}),\label{A3} \\
&& \tA^8_{\mu}=\dfrac{1}{\sqrt 3}(A^1_{\mu}+A^4_{\mu} +A^6_{\mu} ), \label{A8} \\
 &&\tA^I_{\pm\mu}=\dfrac{1}{2}(\tA^1_\mu \pm i \tA^2_\mu),\nn \\
   &&\tA^U_{\pm\mu}=\dfrac{1}{2}(\tA^4_\mu \pm i \tA^5_\mu),\nn \\
     &&\tA^V_{\pm\mu}=\dfrac{1}{2}(\tA^6_\mu \pm i \tA^7_\mu),
     \eea
     where real components of the gluon field $\tA^{1,2,4,5,6,7}_\mu$ are given as follows
     \bea
&& \tA^1_\mu=\dfrac{1}{3 \sqrt 2}(-3A^3_\mu-2 A^1_\mu+\sqrt 3 A^8_\mu+4 A^4_\mu-2 A^6_\mu), \nn\\
&&\tA^2_\mu=-\dfrac{1}{\sqrt 6} (A^3_\mu-2 A^1_\mu+\sqrt 3 A^8_\mu+2 A^6_\mu), \nn \\
&& \tA^4_\mu=\dfrac{1}{3\sqrt 2}(2 \sqrt 3 A^2_\mu-2 A^1_\mu+\sqrt 3 A^5_\mu-\sqrt 3 A^7_\mu-\nn \\
&&~~~~~~~~~~~~~~~2\sqrt 3 A^8_\mu+A^4_\mu+A^6_\mu),\nn\\
&& \tA^5_\mu=- \dfrac{1}{\sqrt 6}(2 A^3_\mu+\sqrt 3 A^5_\mu+\sqrt 3 A^7_\mu+A^4_\mu-A^6_\mu),\nn \\
&& \tA^6_\mu= \dfrac{1}{3\sqrt 6}(-6A^2_\mu-2\sqrt 3 A^1_\mu-3 A^5_\mu+3 A^7_\mu-\nn \\
&&~~~~~~~~~~~~~~~~~~6 A^8_\mu+\sqrt 3(A^4_\mu+A^6_\mu)),\nn \\
&&\tA^7_\mu=\dfrac{1}{3\sqrt 2}(2 \sqrt 3 A^3_\mu-3 A^5_\mu-3 A^7_\mu+\sqrt 3(A^4_\mu-A^6_\mu)).\nn\\
                           \label{A124567}
                           \eea
Expression of the QCD Lagrangian in the Cartan-Weyl basis in terms of new gluon field variables $\tA_\mu^a$
has a general Weyl symmetric multiplet structure providing multi-particle quark-gluon solutions.
One can verify, that the following constraints
\bea
&&A^2_\mu=-A^5_\mu=A^7_\mu\equiv \tC_{3\mu}, \nn \\
 && A^1_\mu=A^4_\mu=A^6_\mu\equiv \tC_{8\mu} , \label{twosinglets}
 \eea
 define  Abelian projection with two gluon singlet fields $\tC_{3\mu}, \tC_{8\mu}$ which coincide
 with Abelian fields $\tC_{3\mu}, \tC_{8\mu}$ given in Section 4.
  \bea
 && \tA_{3\mu}= \sqrt 3 \tC_{3\mu}, \nn \\
 && \tA_{8\mu}=\sqrt 3 \tC_{8\mu}.
 \eea
 The factor $\sqrt 3$ is consistent with three times larger contribution of the Abelian singlet gluon $\tA_8$ to
 one-loop effective action obtained in \cite{flyvb}. Note that only one singlet gluon, $\tA_8$, interacts with the
 singlet quark , due to equations (\ref{eqC3}-\ref{psys2}). This simplifies significantly the calculation of one-loop 
 effective action due to the absence of cross interaction terms between external quark and gluon fields.

Certainly, one can define other Cartan subalgebras containing generator $H_8$ and the second generator
$H'_3$ different from $H_3$, (\ref{H3prim}), for instance, given by the real matrix
 \bea
    &&H'_3=\dfrac{1}{6}\begin{pmatrix}
   2&-1& -1\\
  -1 &-1&2\\
   -1&2&-1\\
 \end{pmatrix}.
 \eea
With these Cartan generators one can reconstruct a full Cartan-Weyl basis,
 however, the obtained real Cartan-Weyl basis will not
 provide two Weyl singlet Abelian fields $\tC_{3\mu}, \tC_{8\mu}$ corresponding to
 generators $H_8, H'_3$ in a consistent manner.
 So that, the complex Cartan-Weyl basis consistent with the Abelian projection with
 two independent singlet Abelian fields is unique, and it provides uniquely the
 equations  (\ref{eqC3}-\ref{psys2})  for the constituent quark.

 \subsection {Manifestation of quark multiplet structure}

 A natural question arises on whether a singlet constituent quark can be observed.
 Certainly, such quarks and single gluons might be observed in the deconfinement phase,
 in the quark-gluon plasma.
 In the confinement phase the color symmetry is preserved, so that the existence
 of singlet constituent quark solutions does not mean that such quarks can be observed
 directly.
 The observable quantities must be gauge invariant under full $SU(3)$ group transformations.
 Due to the presence of three main standard Weyl group representations $\Gamma_N$ ($N=1,2,3$)
 one can construct three different quadratic gauge invariant quark operators
  $\bar {\hat \Psi} \hat \Psi$  composed from the quark multiplets belonging to $\Gamma_N$.
  So that, it is not necessary to perform summation over colors postulated in the conventional
  QCD. An operator $\bar {\hat \Psi} \Psi$ made of singlet quark $\hat \Psi=\psi^s \hu^s$
 describes the simplest heavy meson, charmonium, consisting from one c-quark and one
 c-antiquark, giving an example of indirect manifestation of the existence of a singlet
 constituent quark. In a similar manner, there are three types of vacuum quark condensates
 corresponding to quark multiplets $\Gamma_{1,2,3}$. Such condensates will provide different
 contribution to vacuum energy through the generation
 of a non-trivial quantum effective potential \cite{plb2023}. Note, that equations for the singlet constituent quark
 admits a spectrum of solutions. This implies a corresponding spectrum of the lowest states of the charmonium,
 which can be compared with observational experimental data.
\begin{figure}[h!]
\centering
{\includegraphics[width=80mm,height=55mm]{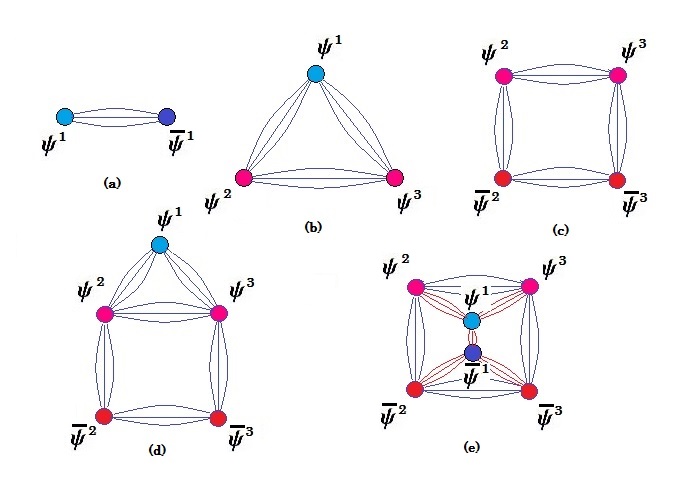}}
\caption[fig1]{Hadron compositions corresponding to Weyl group representations:
(a) $\Gamma_1 \otimes {\bar \Gamma}_1$;
(b)$ \Gamma_3$; (c) $ \Gamma_2\otimes {\bar \Gamma}_2$; (d)$ \Gamma_3 \otimes \Gamma_2$;
(e) $\Gamma_3 \otimes {\bar \Gamma}_3$. Attraction between quarks $\psi^1$ and $\psi^{2,3}$ is
provided by gluon field $A_\mu^8$, and attraction between quarks $\psi^2, \psi^3$ is due to
dominant interaction of gluon field $A_\mu^3$ to compare with $A_\mu^8$, as follows from Eqns.
 (\ref{psys1}).}\label{Fig 1}
\end{figure}
 
Note, that experimental evidences of the existence of eight color gluons and three color quarks
confirm these numbers of the color quarks and gluons as off-shell particles, i.e. virtual particles.
The number of quarks and gluons as real particles is determined by the number of independent 
dynamical single-particle solutions, corresponding to singlet representation $\Gamma_1$ of the Weyl group.
So that, one has only two Abelian gluon singlets $\tC_{3\mu}, \tC_{8\mu}$, as it is shown in section 4.
As for the quarks, one has only one singlet constituent quark $\psi^1\in \Gamma_1$ dressed in a singlet gluon
field $\tC_{8\mu}\in \Gamma_1$.
The remaining two quarks $(\psi^2, \psi^3)$ and gluons $\tC_{3\mu}, \tC_{8\mu}$ satisfy to a reduced system of coupled
equations obtained from Eqs. (\ref{psys1}) by imposing a constraint $\psi^1=0$.  So that, the reduced equations
describe two constituent quarks forming two-dimensional representation $\Gamma_2$. 
It is interesting to note, G. Zweig in the original work \cite{Zweig} introduces 
three color quarks defined by three aces which break up into one isospin quark doublet and one singlet. 

Certainly, the quark and gluon operators
$\hat \Psi, \hat A_\mu$ describe one-particle quantum states which are not Lorentz and gauge invariant,
and they can not be observed directly. It is clear, that one should take into account the
interaction of quark and gluon with vacuum quark and gluon condensates which leads to formation
of observable hadrons. In a case of pure gluodynamics, it has been shown that
such interaction leads to formation of localized bound states corresponding to lightest pure
glueballs \cite{plb2023}. In a similar manner we expect that interaction
of the constituent quark to vacuum gluon condensate will produce  hadrons describing
mixed states of quarks and gluons. This problem will be considered in a separate paper \cite{diez1}.

 The existence of constituent quark multiplets corresponding to Weyl representations $\Gamma_N$
implies that old standard quark model should be modified.
Note that interaction coupling constants of quarks $\psi^p$ with Abelian
gluons $(\tC_{3\mu}, \tC_{8\mu})$ in (\ref{psys1})
are determined by eigenvalues of color charge matrices $L,Q$, (\ref{LQ}).
A simple analysis of Eqs. (\ref{psys1})
shows that due to a special quark-gluon interaction structure of the equations (\ref{psys1}), all three quarks
have mutual attraction. This provides qualitative description of possible hadrons composed from quark Weyl multiplets
$\Gamma_{1,2,3}$, as shown in Fig. 1.
The existence of  singlet quarks allow to construct meson states from one quark and one anti-quark
without need to use $SU(3)$ singlet $\bar\psi^i \psi_i$ with summation over colors.
Some baryons, like $\Omega^-$ hyperon, can be composed from three quarks forming
Weyl representation $\Gamma_3$,
since the total color charge of three p-quarks vanishes, as it  follows from Eqns.  (\ref{psys1}).
  It is clear, that the existence of irreducible representations $\Gamma_{1,2}$
allows the exotic mesons (tetraquarks, pentaquarks etc) to be naturally included in the modified quark model, Fig.1c,d,e.\\

\subsection{Conclusion}

It has been demonstrated that after removing all unphysical pure gauge field degrees of freedom
one can select invariant subspaces of quark and gluon solutions which possess a finite color symmetry
with respect to the Weyl group.
It is really remarkable, that Weyl group symmetry reveals an intrinsic
color symmetry of quarks and gluons, which  leads to deep implications:
(i) the existence of the singlet constituent quark and singlet gluons
allow to introduce strict concepts of quarks and gluons as fundamental particles;
(ii) the existence of a non-degenerate color singlet vacuum which is a necessary condition for the color
confinement phenomenon in QCD in agreement with the observation of Wilson, Kogut and Susskind
\cite{wilson1974, kogut1975}. This implies that the origin of the
color confinement is encoded in the rich group structure of the Yang-Mills theory.
Certainly, the Weyl group symmetry can play an important
role in description of hadron structure and other phenomena in QCD.

\acknowledgments

One of authors (DGP) acknowledges  A.V. Kotikov, A.B. Voitkiv, M.A. Ivanov, L.-P. Zou, N. Korchagin for
valuable discussions, and thanks Prof. I.V. Anikin for invitation to work in BLTP JINR and for numerous 
inspiring discussions.
 This  work is supported by Chinese Academy of Sciences
(PIFI Grant No. 2019VMA0035), National Natural Science Foundation of China (Grant No. 11575254),
and by Japan Society for Promotion of Science (Grant No. L19559).\\

\vskip 1 cm
\end{document}